\documentclass[12pt]{article}

\usepackage{dcolumn}
\usepackage{bm}
\usepackage[latin1]{inputenc}
\usepackage[spanish,english]{babel}
\usepackage{amsfonts}
\usepackage{amssymb}
\usepackage{graphicx}
 \usepackage{setspace}
 \usepackage{verbatim} 
 \usepackage[normalem]{ulem}
 
\usepackage{color}
\usepackage{soul}

 \usepackage{Preamble}

\usepackage[margin=3cm]{geometry}

\newcommand{\be}{\begin{equation}}
\newcommand{\ee}{\end{equation}}
\newcommand{\bea}{\begin{eqnarray}}
\newcommand{\eea}{\end{eqnarray}}

\begin{document}

\begin{titlepage}

\begin{center}
{\large \bf Cosmological mirror symmetry and gravitational-wave helicity} 
\end{center}

\begin{center}

Juan Calder\'{o}n~Bustillo\footnote{juan.calderon.bustillo@usc.es}\\
{\footnotesize \noindent {\it {Instituto Galego de F\'{i}sica de Altas Enerx\'{i}as, Universidade de Santiago de Compostela, 15782 Santiago de Compostela, Galicia, Spain.\\ Department of Physics, The Chinese University of Hong Kong, Shatin, N.T., Hong Kong}}   }\\

Adrian del Rio\footnote{adrdelri@math.uc3m.es}\\ 
{\footnotesize \noindent {\it {Departamento de Matem\'{a}ticas, Universidad Carlos III de Madrid. Avda. de la Universidad 30, 28911 Legan\'{e}s, Spain.}}   }\\

{Samson H. W. Leong}\footnote{samson32081@gmail.com}\\
{\footnotesize \noindent {\it {Department of Physics, The Chinese University of Hong Kong, Shatin, N.T., Hong Kong}}   }\\

Nicolas Sanchis-Gual\footnote{nicolas.sanchis@uv.es}\\
{\footnotesize \noindent {\it {Departamento de Astronom\'{i}a y Astrof\'{i}sica, Universitat de Val\`{e}ncia. Dr. Moliner 50, 46100, Burjassot (Val\`{e}ncia), Spain}}   }\\

\end{center}

\begin{abstract}

Our current understanding of the Universe relies on the hypothesis that, when observed at sufficiently large scales, it looks statistically identical regardless of location or direction of observation.  Consequently, the Universe should exhibit  mirror-reflection symmetry. In this essay, we show that gravitational-wave astronomy provides a unique, observer-independent test of this hypothesis. In particular, we analyze  the average   circular polarization emitted by an ensemble of binary black hole  mergers  detected by LIGO-Virgo,  which we compute using a novel geometric and chiral observable in  general relativity.  We discuss current results and future prospects with upcoming detections and  technical advancements. Moreover, we show that this circular polarization and the helicity of the remnant black hole are linearly correlated, drawing a conceptual parallel with Wu experiment  in particle physics.
 
\end{abstract}

\vspace{0.8cm}

\begin{center}{\it Essay written for the Gravity Research Foundation 2025 Awards for Essays on Gravitation} \end{center}
\begin{center}Submitted on March 31, 2025 \end{center}
\end{titlepage}

The Lambda Cold Dark Matter ($\Lambda$CDM)  model  in Cosmology has been successful in accurately describing a broad collection of cosmological data over decades within a unified framework \cite{Dodelson:2024ft}. This is a simple model  based on a few fundamental assumptions, such as the theory of General Relativity (GR) and the Cosmological Principle. The first hypothesis  states that the expansion of the Universe  is governed by Einstein's gravitational equations, while the second assumes that the matter content is distributed, on average,  homogeneously and isotropically at sufficiently large scales. When  combined, these assumptions imply that the average geometry of our Universe should be described by the Friedmann-Lemaitre-Robertson-Walker (FLRW) metric, written  in suitable  coordinates as $ds^2=-dt^2+a^2(t)d\Sigma^2$, with $a(t)$ describing the expansion of the spatial volume $\Sigma$ over time $t$. Observations of the large-scale galaxy distribution and  the anisotropies in the Cosmic Microwave Background, for example, can be explained by analyzing small perturbations on this spacetime background, provided  additional parameters are introduced to account for dark matter and dark energy. 

Although the $\Lambda$CDM model has been extensively validated over the past decades  with increasingly precise cosmological data,   various  findings in recent years have  challenged its validity or some of its underlying assumptions. These include  discrepancies  in the determination of the Hubble constant by independent cosmological probes, tensions in observational constraints on the strength of matter clustering,  and  other  less statistically  significant anomalies \cite{Abdalla:2022yfr}. While these inconsistencies could stem from unknown systematic errors, there is growing concern that a fundamental issue may be at play, including the possibility that the Cosmological Principle may not strictly hold  \cite{Aluri:2022hzs}. These issues call for novel and complementary cosmological tests.

Since 2015 the  detection of gravitational waves (GWs) emitted by  binary black hole mergers (BBHs) has become a reality, paving the way for examining with increasing precision not only questions of astrophysical nature,  but also fundamental aspects of gravity \cite{PhysRevX.9.031040, PhysRevX.11.021053, PhysRevX.13.041039}.  After a decade of observations and about  100 detected events, we  now  have some potential to  study  global properties of our Universe, such as the analysis of large-scale homogeneity and isotropy. Interestingly, the observed BBHs extend up to distances of 10 Gpc, offering valuable probes of large-scale phenomena. Some observer-dependent tests of homogeneity and isotropy have already been proposed, focusing on  extrinsic source parameters such as  sky location \cite{Essick2023_distributions} or source orientation  relative to the observer \cite{Isi_2024,vitale2022orientationsbinaryblackholes}. In this essay, we describe an independent probe of the large-scale mirror symmetry (and, consequently, of the Cosmological Principle) by targeting instead the handedness or helicity of GWs detected by LIGO-Virgo \cite{CalderonBustillo:2024akj}. This observable directly measures an intrinsic property uniquely determined by the masses and spins of the BBHs, thereby making this test observer-independent.  

Each BBH  can be regarded as a physical system with a natural notion of helicity. As is well known, the remnant black hole typically acquires a recoil velocity. If the  black hole formed after the merger gets ``kicked''   with a linear velocity $\vec v$ and rotates with angular velocity $\vec \Omega$, both measured in a cosmological frame co-moving with the expansion of the Universe, then this black hole will propagate in space with some degree of helicity,  determined by the product $h\sim \vec v\cdot \vec \Omega$,  which is furthermore  constant in time. As a result, we may  classify BBH systems as right-handed if $h>0$ or left-handed if $h<0$,  i.e. depending on whether the resulting final black hole rotates left- or right-handedly along its trajectory. If the cosmological principle truly holds then we expect, on general grounds, that our Universe should contain, {\it on average}, equal proportions of right- vs left-handed BBHs. Otherwise, our Universe would exhibit a net, preferred  handedness, violating mirror symmetry. In particular, since the FLRW metric is symmetric under any reflection transformation of the form $\vec x\to -\vec x$ with respect to any reference plane, the FLRW solution of the Einstein's equations wold fail to describe the {\it averaged geometry} of our observable Universe. 

Although GR  is not a chiral gravitational theory, a Universe containing different populations  of right- and left-handed  BBHs   could naturally emerge from various scenarios. This is because initial conditions in the early universe, or specific black hole formation channels in astrophysics, which may favour one particular handedness, may end up producing such imbalance between right- and left-handed binaries over  time. Thus, the assumption that our Universe must contain equal proportions of chiral BBHs is far from obvious even within the framework of GR, and must be  confirmed with observations. A more fundamental  possibility for large-scale mirror asymmetry is that the underlying theory of gravity itself may be  chiral.  To give some examples, some modified theories of gravity introduce birefringence effects, see e.g. \cite{Ng2023}. While the method we propose can be  also carried out in such alternative  theories, in this essay we restrict our analysis to the first possibility, where mirror asymmetries arise from potential imbalances in BBH populations rather than deviations from GR. 

Interestingly, the aforementioned notion of helicity  leaves a distinctive imprint on the GW signal emitted by the binary.  More precisely, as studied in \cite{leong2025gravitationalwavesignaturesmirrorasymmetry}, the value of the parameter $h$ turns out to be linearly correlated with the following geometric integral, which, as we will soon see,  only depends on GW data~\cite{dRetal20, dR21}: 
\bea
V_{{\rm GW}} := \int_{\mathcal J^+} d \mathcal J^+ \epsilon^{abc}N_{ad}D_b N_c^{\ d} 
\label{chernsimons}\, .
\eea
In this integral, $\mathcal J^+$ denotes the future null infinity of the source, a three-dimensional null hypersurface that physically represents the asymptotic points of infinity where all outgoing light-rays propagate to~\cite{PhysRevLett.10.66}. On the other hand, $d {\cal J}^+$ is the 3-volume  on $\mathcal J^+$ (e.g. $d {\cal J}^+={du}\,{d\cos\theta}\, {d\phi}$ in some spherical coordinate frame $(u,\theta,\phi)$, where $u$ is retarded time), $\epsilon^{abc}$ is the totally antisymmetric tensor, $D_a$ is a natural covariant derivative on $\mathcal J^+$~\cite{PhysRevLett.46.573}, and  $N_{ab}$ denotes the Bondi News geometric tensor~\cite{Geroch1977}. The latter  carries physically valuable information about the GW flux emitted by the binary. Namely, if  $\tilde h^+_{\ell m}(\omega)$ and  $\tilde h^\times_{\ell m}(\omega)$ represent, in a {\it generic frame},  the Fourier transforms of the real and imaginary parts of the GW strain multipoles, $h_{\ell m}(u)=h^+_{\ell m}(u)-i h^{\times}_{\ell m}(u)$, expressed in the basis of spin-weight spherical harmonics, $_{-2}Y_{\ell m}(\theta,\phi)$, then it is possible to write the expression above as
\bea
V_{{\rm GW}}&=& \int_{0}^{\infty} d \omega \,\omega^{3} \sum_{\ell=2}^{\infty}\sum_{m=-\ell}^{+\ell} \quad\left[\bigl|\tilde h^{+}_{\ell m}(\omega)-i \tilde h^{\times}_{\ell m}(\omega)|^{2}-|\tilde h^{+}_{\ell m}(\omega)+i \tilde h^{\times}_{\ell m}(\omega)|^{2}\right] \, . \label{VGW} 
\eea
  Since the combinations $\tilde h^{+}_{\ell m}-i \tilde h^{\times}_{\ell m}$ and $\tilde h^{+}_{\ell m}+i \tilde h^{\times}_{\ell m}$ represent  left-  and right-handed  circularly polarized wave modes, respectively\footnote{Equivalently, $\tilde h^{+}_{\ell m}-i \tilde h^{\times}_{\ell m}$ and $\tilde h^{+}_{\ell m}+i \tilde h^{\times}_{\ell m}$ represent the  Fourier modes of fields with  spin-weight $-2$ and $2$, respectively.}, the observable  $V_{\rm GW}$ can be interpreted as the gravitational analogue of the Stokes $V$ parameter in optics,  quantifying the net circular polarization  of the GW flux emitted by the source.
 
\begin{figure}[htb]
    \centering
    \includegraphics[width=0.49\linewidth]{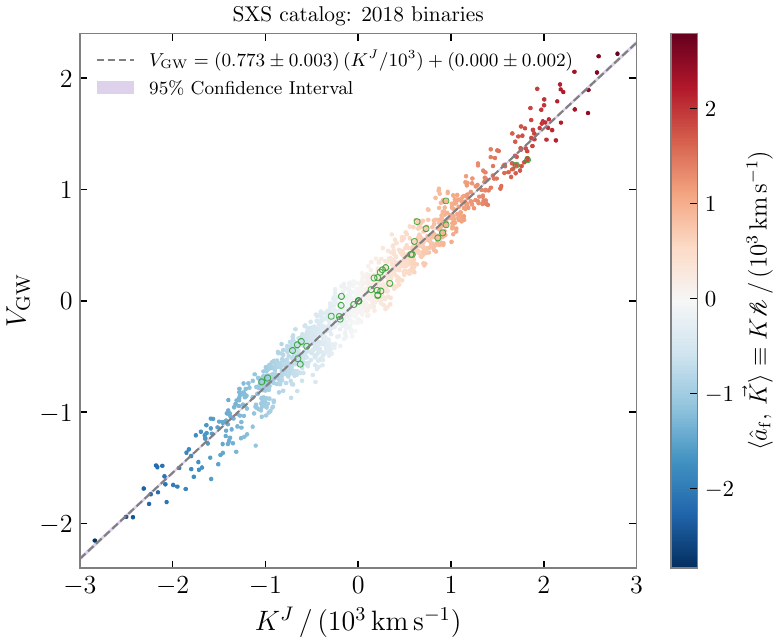}
    \includegraphics[width=0.49\linewidth]{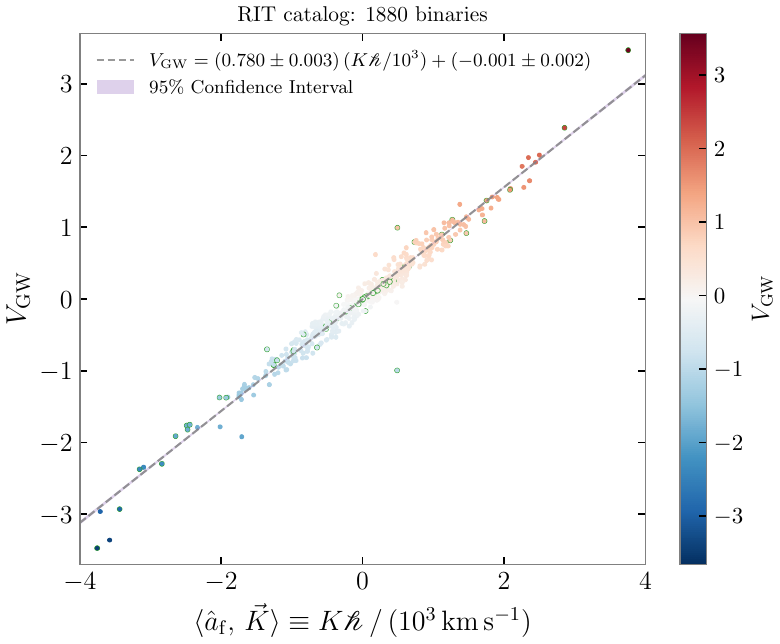}
 \caption{ {\bf Linear correlation found between  GW circular polarization ($V_{\rm GW}$) and the helicity of the final black hole in BBH simulations from the SXS (left) and RIT (right) catalogues.} The helicity  is understood as  the projection of the recoil velocity $\vec {K}$ along either  the total angular momentum direction $\hat J$ or   the final black hole spin $\hat a_{\rm f}$.}
    \label{linearcorrelation}
\end{figure}

   Numerical relativity simulations for quasi-circular and eccentric precessing BBHs from both the SXS and RIT catalogues \cite{RIT2022:4thCatalog, SXS2019:Catalog}, as well as   the state-of-the-art surrogate model for quasi-circular precessing mergers, \texttt{NRSur7sq4}~\cite{islam2023analysisgwtc3fullyprecessing}, confirm  that $V_{{\rm GW}}\sim 10^{-3} \cdot h$ \cite{leong2025gravitationalwavesignaturesmirrorasymmetry}. This linear correlation is nicely ilustrated in Fig. \ref{linearcorrelation}.  
 
 While  the concept of black hole helicity $h$ is physically  intuitive and  useful for gaining insights, $V_{\rm GW}$ is more precise from a mathematical viewpoint\footnote{The  definition of $h$ relies on vector directions in Euclidean space and, therefore, ignores general-relativistic details,  and might depend on the choice of frame, among other factors.}.  For a given astrophysical source, equation (\ref{VGW}) can also be used as an estimator to quantify the degree of mirror asymmetry in the underlying spacetime. This may not be surprising in view of the presence of $\epsilon^{abc}$ in  (\ref{chernsimons}),  an expression which resembles a Chern-Simons current in gauge theories, and makes $V_{\rm GW}$ a chiral observable. In particular, using simple arguments, one can infer that mirror symmetry necessarily implies $V_{GW}=0$, i.e. no circular polarization.  To see this, notice that a BBH system with mirror symmetry, as is the case when the  two rotating black holes have parallel spins,  always admits a frame where the conditions $\tilde h^+_{\ell \, -m}=(-1)^\ell  \tilde h^+_{\ell m}$ and $\tilde h^{\times}_{\ell \, -m}=-(-1)^\ell  \tilde h^{\times}_{\ell m}$ are satisfied. This can be seen from the properties of the  spin-weight spherical harmonics, $_{-2}Y_{\ell m}(\theta,\phi)$. Consequently, although a specific GW mode $(\tilde h^{+}_{\ell m},\tilde h^{\times}_{\ell m})$ may exhibit  circular polarization, in the sense of  $|\tilde h^+_{\ell m}-i \tilde h^{\times}_{\ell m}|^2-|\tilde h^{+}_{\ell m}+i \tilde h_{\ell m}^{\times}|^2\neq 0$, its opposite mode $(\tilde h^+_{\ell \, -m},\tilde h^{\times}_{\ell\, -m})$ cancels this contribution  in the  sum over $m$ in  (\ref{VGW}). Physically, if this astrophysical source  emits a circularly polarized flux in a given direction of the sky, an observer in the opposite direction will receive exactly the same flux but with opposite circular polarization.  The net contribution, when integrated over the full celestial sphere, results in zero  handedness. As a result, an imbalance  between left- and right-handed GW modes in Eq.~(\ref{VGW}) can only arise in spacetimes with  mirror asymmetry. 

Single astrophysical sources can indeed produce an excess of  left-handed over right-handed GWs in our Universe, or viceversa, resulting in $V_{\rm GW}\neq 0$. Since $V_{\rm GW}$ changes sign under a mirror transformation (right-handed modes become left-handed modes and viceversa), these sources break mirror symmetry, as explained above. Nevertheless, if the Cosmological Principle holds, mirror-reflection symmetry should be restored at sufficiently large scales when averaging across all GW observations in our Universe, meaning that we should obtain $\langle V_{\rm GW} \rangle= 0$. Consequently, GW catalogs \cite{PhysRevX.9.031040, PhysRevX.11.021053, PhysRevX.13.041039} can be used to constrain mirror asymmetry in our Universe through the observable $V_{{\rm GW}}$.

Interestingly, since $V_{{\rm GW}}$ can be expressed as the curvature integral (\ref{chernsimons}), it remains invariant under any coordinate transformation that preserves the  handedness of the reference frame. In other words, $V_{{\rm GW}}$ is a (dimensionless) real number  intrinsic to the astrophysical source, independent of the chosen frame and  solely dependent on physical GW data, including its intrinsic handedness. Unlike previous works  \cite{Essick2023_distributions,Isi_2024,vitale2022orientationsbinaryblackholes}, this enables  an examination of the Cosmological Principle that is independent of the source's orientation with respect to the observer.

{\bf Mirror asymmetry in single LIGO-Virgo detections.} We now estimate  $V_{\rm GW}$ for each event within a fully Bayesian framework, which accounts for parameter uncertainties arising from both the limited signal loudness  (compared to the detector noise) and parameter degeneracies. Given the GW data $d$ for each event, our signal model $h(\vec\theta)$ for source parameters $\vec\theta$, and the a-priori probability distribution $p(\vec\theta)$, we obtain posterior probabilities for the source parameters $p(\vec\theta | d)$. With this, we can derive the posterior distribution for $V_{\rm GW}$ and the corresponding prior, $p(V_{\rm GW}|d)$ and $p(V_{\rm GW})$, via Eq.~\ref{VGW}.  If the data is informative about $V_{\rm GW}$ then the posterior will deviate from the prior. Otherwise, the posterior will simply follow the prior. %

\begin{figure}[t!]
\begin{center}
\includegraphics[width=0.49\textwidth]{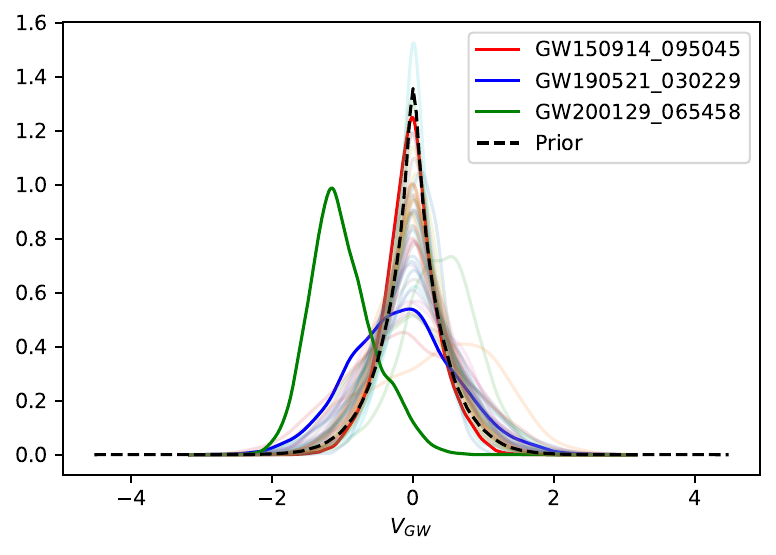}
\includegraphics[width=0.49\textwidth]{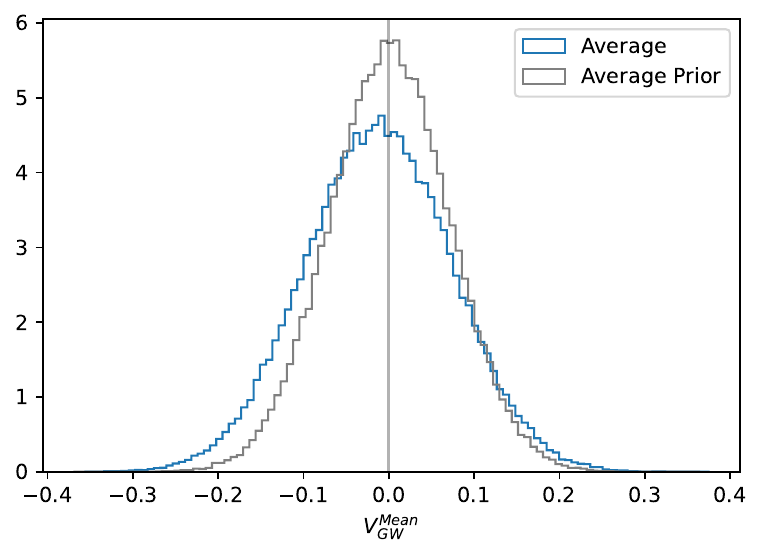}
\caption{\textbf{Left panel: Posterior probability distribution of the gravitational Stokes parameter  (\ref{VGW}) for the 47 LIGO-Virgo events considered  in~\cite{islam2023analysisgwtc3fullyprecessing}}. The black curve denotes the prior distribution.  Only GW200129 (in green) shows strong evidence for a nonzero $V_{\rm GW}$.  \textbf{Right panel: Posterior probability distribution (in blue) for the average value of $V_{\rm GW}$ across the 47 LIGO-Virgo events.}   The  prior distribution is shown in gray. While minimally informative, the posterior remains strongly consistent with zero. \vspace{1cm}} 
\label{asymmetry}
\end{center}
\end{figure}

 Using the posterior distributions for the masses and spins of 47 BBH mergers  obtained by~\cite{islam2023analysisgwtc3fullyprecessing}, we  calculate the posterior distributions for $V_{\rm GW}$ via Eq.~\ref{VGW}, generating the GW modes $h_{\ell m}$ (up to $\ell=4$) using the surrogate \texttt{NRSur7dq4} model\footnote{For BBHs of total mass $M$, the GW modes decay fast with increasing $\ell$ and peak around $\omega\sim M^{-1}$ in frequency domain. Thus, the infinite sum and integral of (\ref{VGW}) can be well approximated with $\ell\leq 4$ and the relevant frequency window  \cite{CalderonBustillo:2024akj}.}. Figure~\ref{asymmetry} displays the obtained posterior probability distribution $p(V_{\rm GW}|d)$ for all the 47 events, along with the corresponding prior $p(V_{\rm GW})$, plotted in black and peaking around zero. Among all events,  only GW200219 (in green) exhibits a posterior distribution  that significantly deviates from zero, peaking around $V_{\rm GW} = -1$, and with $V_{\rm GW} = 0$ laying at the $98th$ percentile. 

\begin{figure*}[t!]
\begin{center}
\includegraphics[width=1\textwidth]{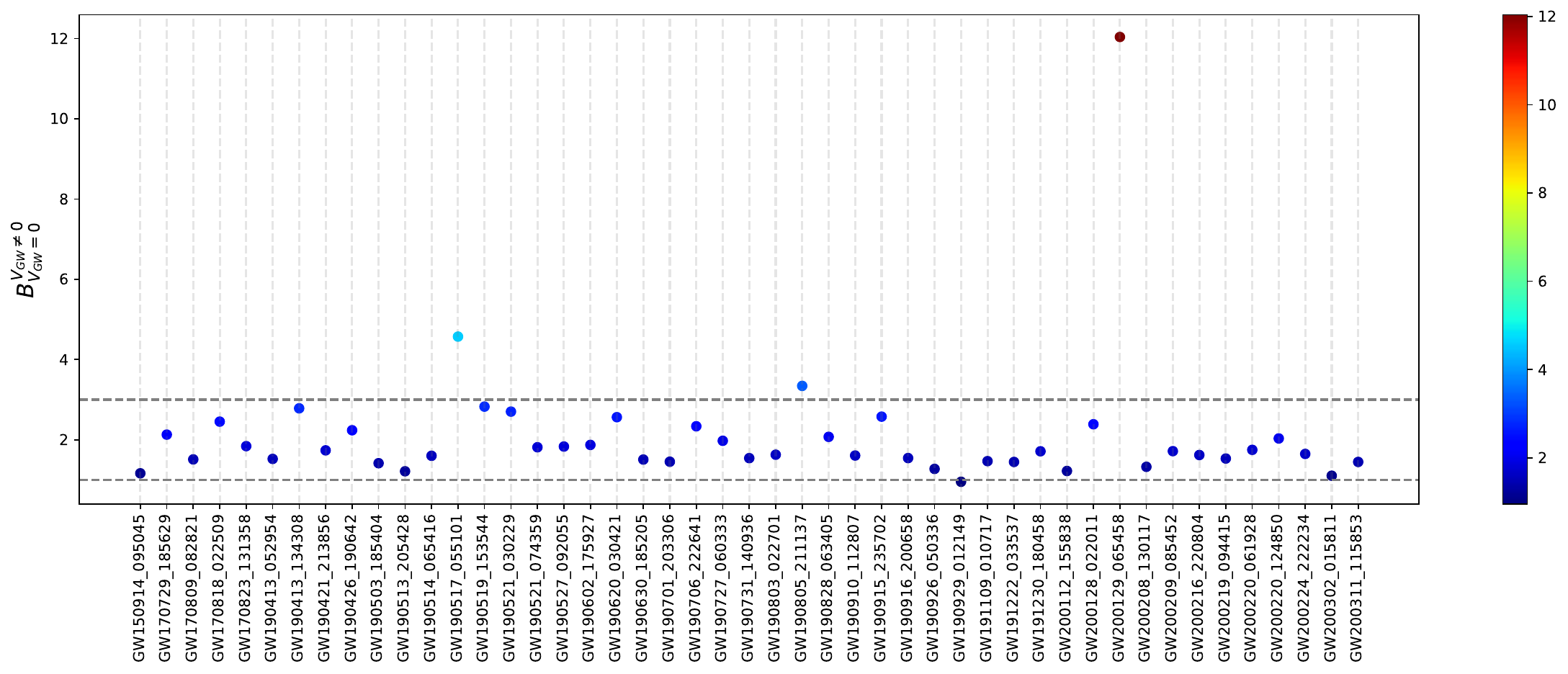}
\caption{\textbf{Evidence for non-zero $V_{\rm GW}$ in each analyzed event}. The plot shows the relative Bayes factor comparing the $V_{\rm GW}\neq 0$ scenario against the $V_{\rm GW} = 0$ one, based on the posterior parameter samples released by~\cite{islam2023analysisgwtc3fullyprecessing}.}
\label{allinarow}
\end{center}
\end{figure*}

We can further quantify the  evidence for $V_{\rm GW}\neq 0$ for each event using the Savage-Dickey ratio, defined as ${\cal{B}}^{V_{\rm GW}\neq 0}_{V_{\rm GW}=0} := p(0)/p(0|d)$. A higher value indicates stronger statistical evidence  in favor of $V_{\rm GW}\neq 0$ against the null hypothesis $V_{\rm GW}=0$. The resulting values in our study are illustrated in Fig.~\ref{allinarow}. As before, only one  case (GW200219) stands out  with strong evidence for $V_{\rm GW} \neq 0$, with ${\cal{B}}^{V_{\rm GW}\neq 0}_{V_{\rm GW}=0} = p(0)/p(0|d) = 12.7$ or, equivalently, $92.7\%$ probability. By performing our own parameter inference, we obtain a similar result of ${\cal{B}}^{V_{\rm GW}\neq 0}_{V_{\rm GW}=0} = 16.7$, corresponding to a $94.3\%$ probability \cite{CalderonBustillo:2024akj}.

As argued in \cite{dRetal20}, BBHs emitting $V_{\rm GW}\neq 0$   require precessing dynamics (non-precessing BBHs are mirror-symmetric relative to the orbital plane). The challenges found above for identifying BBHs with net circular polarization are connected to the current limitations of  waveform modeling for binaries with orbital precession and, more importantly, to the fact that existing matched-filter searches for GWs {\it only} target aligned-spin systems (which are non-precessing).  Future improvements in  waveform modelling and search techniques will   enhance the identification of events with $V_{\rm GW}\neq 0$.

{\bf Ensemble properties: average mirror symmetry.} The right panel of Figure~\ref{asymmetry} shows the posterior distribution for the average value of $V_{\rm GW}$ across all 47 events (in blue) alongside the corresponding prior. The posterior distribution is consistent with zero,  deviating only mildly from the prior. More precisely, our results yield a median value of $\langle V_{\rm GW} \rangle= -0.013^{+0.142}_{-0.141}$ within the symmetric $90\%$ credible interval. Consistently, applying the Savage-Dickey ratio  again, we find no significant evidence of mirror symmetry violation, obtaining a relative evidence of 1.27:1 in favor of the $\langle V_{\rm GW} \rangle\neq 0$ hypothesis over   $\langle V_{\rm GW} \rangle=0$. Additionally, we  verify that removing GW200129 from the dataset causes the posterior to closely follow the prior.

Given the one-to-one correlation between black hole helicity and GW circular polarization discussed earlier (recall Fig \ref{linearcorrelation}), the experiment of measuring (\ref{VGW}) across all GW detections can be understood  as the gravitational analogue of Wu's experiment~\cite{PhysRev.105.1413}. That experiment investigated  parity symmetry breaking in weak interactions, which governs the spontaneous decay of particles. Namely, by analyzing the beta decay of an ensemble of Cobalt atoms---focusing on the distribution of emitted linear momentum  relative to the atom's spin (or, equivalently, the helicity of the emitted electrons)---a preferred direction of emission was observed, revealing  a violation of mirror symmetry in weak interactions. A conceptually similar procedure is proposed in this essay to test the large-scale mirror symmetry. Here, the recoil of the remnant black hole serves the same role of the emitted electron's linear momentum. Thus, by measuring the distribution of remnant black hole linear momenta relative to their own spin, i.e. the helicity of the ``emitted'' black hole after the merger, it becomes possible to probe the large-scale mirror (a)symmetry in our Universe. 

{\bf Future prospects.} Although our results are consistent with the Cosmological Principle, they are far from being conclusive due to  current technical limitations in waveform modeling and matched filter searches, as well as the relatively small dataset used in this study.
Forthcoming BBH observations and theoretical advancements in the coming decades will enable us to make stronger claims regarding  whether our Universe exhibits a preferred handedness, especially when we can  reproduce the same data analysis  with  thousands of available GW detections. 

With such a large dataset, if we can further associate   each value of $V_{\rm GW}$ with its corresponding BBH  location in space with good precision, we will be able to fill in each point on a celestial sphere with all observed values of $V_{\rm GW}$. This will allow us to study spatial correlations, such as the two-point function $\langle V_{\rm GW}(x) V_{\rm GW}(y) \rangle$, as well as to examine fluctuations around the average value $\langle V_{\rm GW}\rangle(=0)$, similar to the standard analysis of temperature anisotropies in the Cosmic Microwave Background. This complementary analysis would provide far more information than the simple space-average values $\langle V_{\rm GW}\rangle$ presented in this essay.

\begin{figure*}[t!]
\begin{center}
\includegraphics[width=1\textwidth]{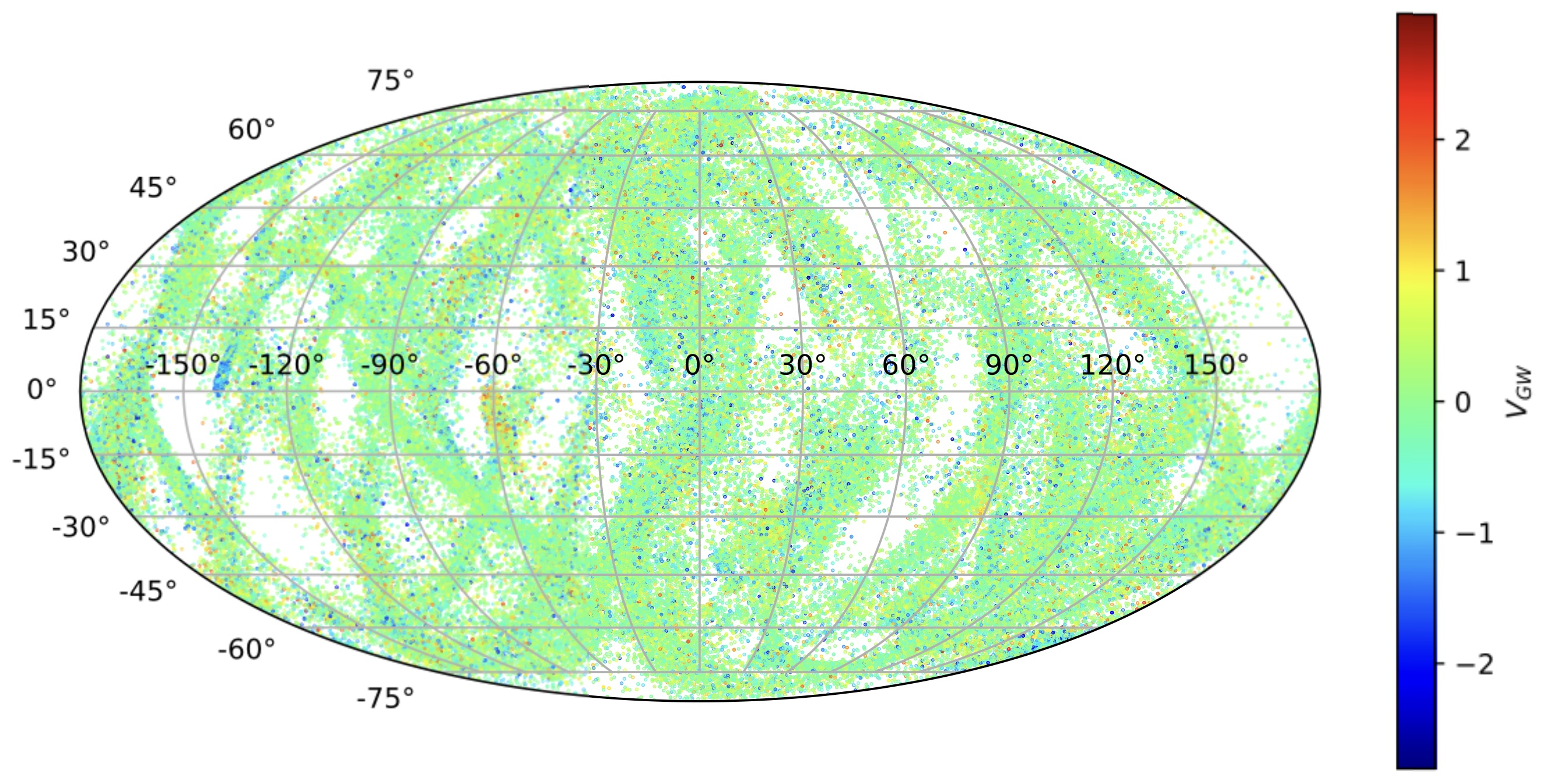}
\caption{ Value of $V_{\rm GW}$ as a function of  sky location for the 47 BBHs analyzed in~\cite{islam2023analysisgwtc3fullyprecessing}. The density of points is proportional to the joint posterior probability for a given sky-location across the 47 BBHs, while the color represents the corresponding   $V_{\rm GW}$ value.\vspace{1cm}}
\label{cmb}
\end{center}
\end{figure*}

To illustrate this idea, a preliminar map using current GW data is displayed in Fig. \ref{cmb}. The sparsity of  points is due to current uncertainties in the sky location of the GW events. In particular, the density of points is proportional to the posterior probability for the sky location across events, with the color denoting the value of  $V_{\rm GW}$. For instance, the clear blue region in the left hand side of the equatorial line in the plot corresponds to the event GW200129, which is one of the best localized events in the third observing run and the only one showing a clear non-zero $V_{\rm GW}$.

Finally, analyzing other GW signals, such as the recently detected stochastic GW background \cite{Agazie_2023},  will also be valuable for probing the large-scale mirror symmetry in other frequency bands.

\newpage

\noindent{\bf{\em Acknowledgments.}}
%
 JCB is funded by a fellowship from ``la Caixa'' Foundation (ID100010474) and from the European Union's Horizon2020 research and innovation programme under the Marie Skodowska-Curie grant agreement No 847648. The fellowship code is LCF/BQ/PI20/11760016. JCB is also supported by the research grant PID2020-118635GB-I00 from the Spain-Ministerio de Ciencia e Innovaci\'{o}n and by its Ram\'on y Cajal program (grant RYC2022-036203-I). 
ADR acknowledges support through {\it Atraccion de Talento Cesar Nombela} grant No 2023-T1/TEC-29023, funded by Comunidad de Madrid (Spain); as well as   financial support  via the Spanish Grant  PID2023-149560NB-C21, funded by MCIU/AEI/10.13039/501100011033/FEDER, UE.
NSG acknowledges support from  the Spanish Ministry of Science and Innovation via the Ram\'on y Cajal programme (grant RYC2022-037424-I), funded by MCIN/AEI/ 10.13039/501100011033 and by ESF Investing in your future. NSG is further supported by the Spanish Agencia Estatal de Investigaci\'on (Grant PID2021-125485NB-C21) funded by MCIN/AEI/10.13039/501100011033 and ERDF A way of making Europe, and by the European Union's Horizon 2020 research and innovation (RISE) programme H2020-MSCA-RISE-2017 Grant No. FunFiCO-777740 and by the European Horizon Europe staff exchange (SE) programme HORIZON-MSCA2021-SE-01 Grant No. NewFunFiCO-101086251. 
%

\bibliographystyle{utphys}
\bibliography{References}

\end{document}